\documentclass[onecolumn,manuscript,preprint,aps,pra]{revtex4-1}
\usepackage{amssymb}

\usepackage{epsfig}
\usepackage[english]{babel}
\usepackage{latexsym}
\usepackage{graphics}
\usepackage{subfigure}
\usepackage{epsfig}
\usepackage{graphicx}
\usepackage{dcolumn}
\usepackage{amsmath}

\usepackage{graphicx}
\usepackage{dcolumn}
\usepackage{bm}

\begin{document}

\title{Terrace-like structure in the above-threshold ionization spectrum of an atom in an IR+XUV two-color laser field}
\author{Kui Zhang$^{1}$, Jing Chen$^{2,3}$, Panming Fu$^{1}$, Zong-Chao Yan$^{4,5,6}$, and Bingbing Wang$^{1}$}

\email{wbb@aphy.iphy.ac.cn}
\affiliation{$^1$Laboratory of Optical Physics, Beijing National Laboratory for Condensed Matter Physics,
Institute of Physics, Chinese Academy of Sciences, Beijing 100190, China}
\address{$^2$ Key Laboratory of High Energy Density Physics Simulation, CAPT, Peking University, Beijing 100084, China}
\address{$^3$ Institute of Applied Physics and Computational Mathematics, P. O. Box 8009, Beijing 100088, China}
\address{$^4$State Key Laboratory of Magnetic Resonance and Atomic and Molecular Physics, Wuhan Institute of Physics and Mathematics, Chinese Academy of Sciences, Wuhan 430071, China}
\address{$^5$ Center for Cold Atom Physics, Chinese Academy of Sciences, Wuhan 430071, China}
\address{$^6$ Department of Physics, University of New Brunswick,
Fredericton, New Brunswick, Canada E3B 5A3}

\date{\today}

\begin{abstract}
Based on the frequency-domain theory, we investigate the above-threshold ionization (ATI) process of an atom in a two-color laser field with infrared (IR) and extreme ultraviolet (XUV) frequencies, where the photon energy of the XUV laser is close to or larger than the atomic ionization threshold. By using the channel analysis, we find that the two laser fields play different roles in an ionization process, where the XUV laser determines the ionization probability by the photon number that the atom absorbs from it, while the IR laser accelerates the ionized electron and hence widens the electron kinetic energy spectrum. As a result, the ATI spectrum presents a terrace-like structure. By using the saddle-point approximation, we obtain a classical formula which can predict the cutoff of each plateau in the terrace-like ATI spectrum. Furthermore, we find that the difference of the heights between two neighboring plateaus in the terrace-like structure of the ATI spectrum increases as the frequency of the XUV laser increases.

\end{abstract}



\maketitle
\section{introduction}

The physics of above-threshold ionization (ATI) of an atom in a strong laser field has been the subject of intensive study for more than thirty years since its first discovery by Agostini \textsl{et al.}~\cite{a1}. The ATI process shows that the ionized electron can experience free-free transition beyond the absorption of minimum photon number necessary for its ionization from an atom, which has refreshed the photoelectric effect found more than a hundred years ago. Recently, with the application of high-order harmonic generation~\cite{hhg1,hhg2,hhg3} and the development of free-electron laser technology~\cite{fel1, fel2}, the laser-dressed XUV/x-ray photo-ionization~\cite{schins,tong} and photo-emission of an atom become hot topics in strong field physics. Veniard \emph{et al}.~\cite{maquet} demonstrated the laser-assisted effect in atomic ionization process by using high-order harmonic radiation. Furthermore, a sideband plateau in the photoelectron spectrum of an atom in an XUV+IR two-color laser field has been predicted in theory~\cite{fe,m,kk} and realized in experiment~\cite{meyer}.

In this work, we apply the frequency-domain theory to investigate the ATI process of an atom in a two-color laser field with IR and XUV frequencies, where the photon energy of the XUV laser is close to or larger than the atomic ionization threshold. Using the channel analysis, we find that the two laser fields play different roles in an ionization process, where the XUV laser determines the ionization probability by the photon number absorbed by the atom, while the IR laser accelerates the electron and hence extends the electron kinetic energy spectrum. As a result, the ATI spectrum shows a terrace-like structure. These results may be useful in understanding the interaction between matters and high-frequency lasers.

\section{Theoretical method}

Based on the frequency-domain theory of laser-matter interaction~\cite{guods,gaolh,fu,wang1,guoyc,wang2,wang3}, our system of an atom and photons can be treated as an isolated system whose total energy keeps constant during a dynamical transition; hence the formal theory of scattering processes~\cite{18} can be applied. Consider an atom interacting with a two-color linear polarized laser field. The Hamiltonian of this atom-laser system (atomic units are used throughout unless otherwise stated) is $H=H_{0}+U\left(\mathbf{r}\right)+V$, where $H_{0}=\frac{\left(-i\nabla\right)^{2}}{2}+\omega_{1}N_{1}+\omega_{2}N_{2}$ is the energy operator for a free electron-photon laser system, $N_{1}$ and $N_{2}$ are the photon number operators of the two fields with frequency $\omega_1$ and $\omega_2$, respectively; $U({\bf r})$ is the atomic binding potential,
and $V$ is the electron-photon interaction operator.
The transition matrix from the initial state $\psi_i$ to the final state $\psi_f$ of the whole system can be expressed as~\cite{wang1,wang2,guoyc}
\begin{eqnarray}
T_{fi}=\langle\psi_{f}|V|\psi_{i}\rangle+\langle\psi_{f}|U\frac{1}{E_{i}-H+i\varepsilon}V|\psi_{i}\rangle,
\label{e1}
\end{eqnarray}
where the first and second terms correspond to the processes of direct and rescattering ATI, respectively. Therefore, the transition matrix can be simply expressed as $T_{fi}=T_{d}+T_{r}$, where $T_d$ represents the transition of the electron from the initial state $|\psi_{i}\rangle$ to the final state $|\psi_{f}\rangle$ by the interaction $V$ with the laser field, and $T_r$ represents the process where the electron recollides with the core after it is ionized by the laser field from the initial state~\cite{wang1}.

In Eq.~(\ref{e1}), the initial state $|\psi_i \rangle =\Phi_i({\bf r}) \otimes | n_1 \rangle\otimes | n_2 \rangle$ with $\Phi_i({\bf r})$ being the ground state of the atom and $| n_j \rangle$ being the Fock state of the laser mode with the photon number $n_j$ for $j=1, 2$. The total energy of the whole system for the initial state $|\psi_i\rangle$ is
$E_{i}=-I_{p}+\left(n_{1}+\frac{1}{2}\right)\omega_{1}+\left(n_{2}+\frac{1}{2}\right)\omega_{2}$ with $I_p$ being the ionization threshold of the atom. The final state $|\psi_f\rangle=|\Psi_{\mathbf{p}_{f}m_{1},m_{2}}\rangle$ is the quantized-field Volkov state of an electron in a two-color laser field~\cite{guo1992b}
\begin{align}
 \label{e2}
|\Psi_{\mathbf{p}_{f},m_{1},m_{2}}\rangle = & V_{e}^{-1/2}\exp\left[i\left(\mathbf{p}_f+u_{p_{1}}\mathbf{k}_{1}
+u_{p_{2}}\mathbf{k}_{2}\right)\cdot\mathbf{r}\right]\sum_{j_{1}=-m_{1},j_{2}=-m_{2}}^{\infty}
\mathcal{J}_{j_{1},j_{2}}\left(\zeta\right)^{*}\nonumber\\
 & \times\exp\left\{ -i\left[j_{1}\left(\mathbf{k}_{1}\cdot\mathbf{r}
 +\phi_{1}\right)+j_{2}\left(\mathbf{k}_{2}\cdot\mathbf{r}+\phi_{2}\right)\right]\right\} |m_{1}+j_{1},m_{2}+j_{2}\rangle,
\end{align}
where $V_{e}$ is the normalization volume for the wave function of the Volkov state, $\phi_{1}$ and $\phi_{2}$ are the initial phases of the two laser fields, $\mathbf{p}_f$ is the final canonical momentum of the ionized electron, and $\mathbf{k}_{1}$ ($\mathbf{k}_{2}$) is the photon momentum of the first (second) laser field. Moreover, $u_{p_{1}}=\widetilde{U}_{p_{1}}/\omega_{1}$ ($u_{p_{2}}=\widetilde{U}_{p_{2}}/\omega_{2}$) with $\widetilde{U}_{p_{1}}$ ($\widetilde{U}_{p_{2}}$) being the ponderomotive energy of the first (second) laser field. The total energy of the whole system for the final state $|\Psi_{\mathbf{p}_{f},m_{1},m_{2}}\rangle$ is $E_{\mathbf{p}_{f},m_{1},m_{2}}=\mathbf{p}_{f}^{2}/2
+\left(m_{1}+\frac{1}{2}\right)\omega_{1}+\widetilde{U}_{p_{1}}+\left(m_2+\frac{1}{2}\right)\omega_{2}+\widetilde{U}_{p_{2}}$.
In Eq.~(\ref{e2}), $\mathcal{J}_{j_{1},j_{2}}\left(\zeta\right)^{*}$ is the complex conjugate of the generalized Bessel function, which can be written as~\cite{guo1992b}
\begin{equation}
\label{gbessel_linear}
\mathcal{J}_{q_{1},q_{2}}\left(\zeta\right)=\underset{q_{3}q_{4}q_{5}q_{6}}{\sum}J_{-q_{1}+2q_{3}+q_{5}+q_{6}}(\zeta_{1})
J_{-q_{2}+2q_{4}+q_{5}-q_{6}}(\zeta_{2})J_{-q_{3}}(\zeta_{3})J_{-q_{4}}(\zeta_{4})J_{-q_{5}}(\zeta_{5})J_{-q_{6}}(\zeta_{6}).
\end{equation}
Here, $\zeta\equiv\left(\zeta_{1},\zeta_{2},\cdots,\zeta_{6}\right)$ and the arguments of the Bessel functions are
\begin{equation}
 \label{gbesselarguments}
\begin{cases}
\zeta_{1}=2\sqrt{\frac{u_{p_1}}{\omega_1}}|\mathbf{p}_f\cdot\hat{\epsilon}_1|,\:\:  \zeta_{2}=2\sqrt{\frac{u_{p_2}}{\omega_2}}|\mathbf{p}_f\cdot\hat{\epsilon}_2|,\\
\zeta_{3}=\frac{1}{2}u_{p_{1}},\:\:
\zeta_{4}=\frac{1}{2}u_{p_{2}},\\
\zeta_{5}=2\frac{\sqrt{u_{p_{1}}u_{p_{2}}\omega_1\omega_2}}{\omega_{1}+\omega_{2}}, \:\:\zeta_{6}=2\frac{\sqrt{u_{p_{1}}u_{p_{2}}\omega_1\omega_2}}{\omega_{1}-\omega_{2}},
\end{cases}
\end{equation}
where $\hat{\epsilon}_1=\hat{\epsilon}_2$ represents the polarization direction of the two laser fields.

In this work, we consider a two-color laser field to be composed of IR and XUV laser fields. Without loss of generality, we assume that the first laser with frequency $\omega_1$ is the IR laser and the second one with $\omega_2$ is the XUV laser. Because $\omega_2$ is much larger than $\omega_1$, the influence of the the relative phase between these two laser fields on the ATI spectrum can be ignored; hence we set $\phi_1=\phi_2=0$ for simplicity in the following calculation. Additionally,  we notice that, since the contribution of the second term $T_r$ in Eq.~(\ref{e1}) is much smaller than that of the first term $T_d$ under our present computational condition, we only consider the first ATI term here, and will discuss the contributions of the rescattering ATI term in a future work. By using the initial and final states, the direct ATI term can be expressed as~\cite{guo1992b}

\begin{equation}
 \label{Td}
T_{d}=V_{e}^{-1/2}\left(\widetilde{U}_{p_{1}}+\widetilde{U}_{p_{2}}-q_1\omega_1-q_2\omega_2\right)
\Phi_{i}\left(\mathbf{p}_{f}\right)\mathcal{J}_{q_{1},q_{2}}\left(\zeta_{f}\right)
\end{equation}
with $q_1$ ($q_2$) being the photon number that the atom absorbs from the IR (XUV) laser field.

\section{terrace-like structure of ATI spectrum}

\begin{figure}[h!]
\includegraphics[width=14cm, height=10cm]{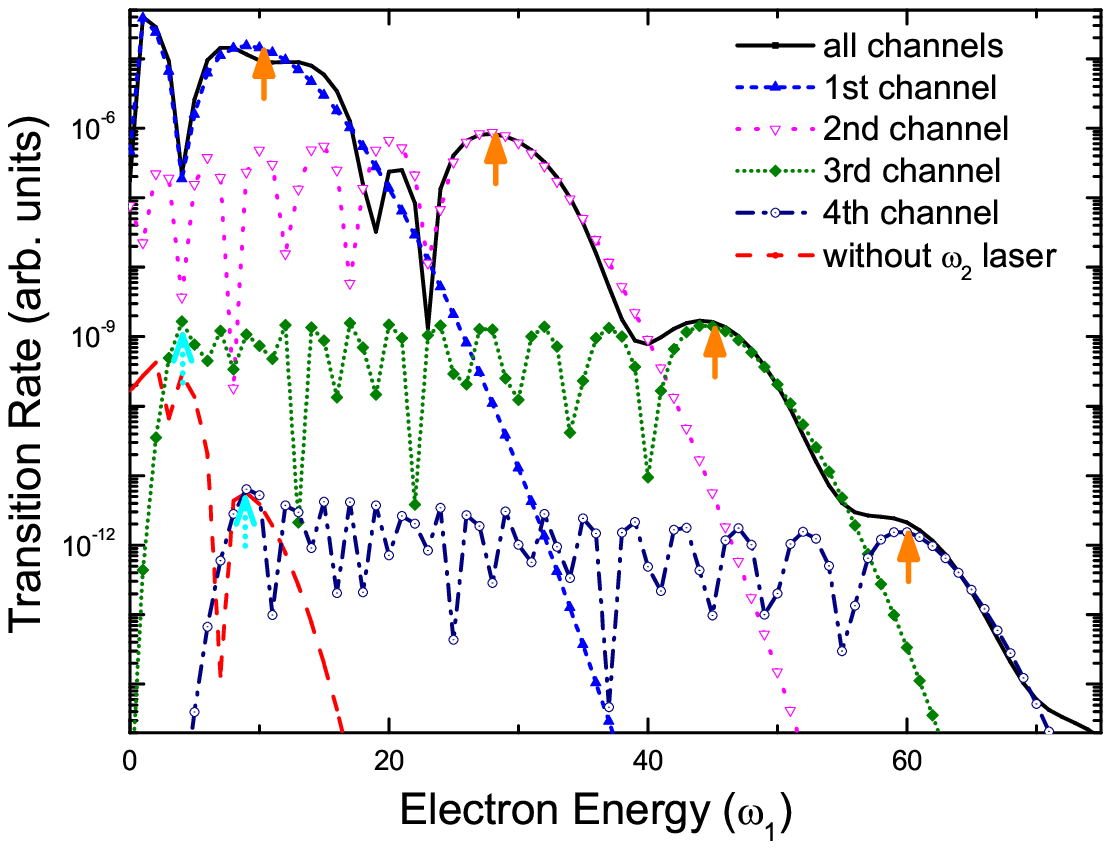}
\caption{(Color online) ATI spectrum of an atom with ionization threshold $I_p=12.1$~eV in a two-color laser field with $\omega_1=1.165$~eV and $\omega_2=10\omega_1$ for laser intensity of $I_1=I_2=3.6\times 10^{13}$ W/cm$^2$.}
\label{f1}
\end{figure}

Figure~\ref{f1} presents the ATI spectra when an atom with ionization threshold $I_p=12.1$ eV interacts with a two-color laser field with the frequencies of $\omega_1=1.165$ eV and $\omega_2=10\omega_1$, and the lasers' intensities are $I_1=I_2=3.6\times10^{13}$ W/cm$^2$ corresponding to $\widetilde{U}_{p_{1}}=3.2\omega_1$ and $\widetilde{U}_{p_{2}}=0.032\omega_1$. Under the above atom-laser conditions, we find that the ATI spectrum is dominated by the direct ATI process in Eq.~(\ref{Td}), and hence the contribution of recollision term is ignored in this work. As shown in Fig.~\ref{f1}, the most remarkable characteristic of the ATI spectrum is the terrace-like structure. To our knowledge, this has not been seen before.

In order to explain the terrace-like structure shown in Fig.~\ref{f1}, we analyze the generalized Bessel function in Eq.~(\ref{Td}) that we believe causes  the formation of this type of structure,\textit{ i.e.} we have $T_d\propto\mathcal{J}_{q_{1},q_{2}}\left(\zeta_{f}\right)$. By using the above laser conditions, we find that the absolute values of the arguments $\zeta_4$, $\zeta_5$, and $\zeta_6$ of the Bessel functions are far less than one, and hence we can make an approximation that  $J_{-q_{i}}(\zeta_{i})\approx 1$ and $q_i=0\left(i=4,5,6\right)$. Therefore, the generalized Bessel function Eq.~(\ref{gbessel_linear}) can be simplified as

\begin{align}
 \label{gbessel}
\mathcal{J}_{q_{1},q_{2}}\left(\zeta\right)\approx & J_{-q_{1}}(\zeta_{1},\zeta_{3})J_{-q_{2}}(\zeta_{2}),
\end{align}
where $J_{-q_{1}}(\zeta_{1},\zeta_{3})=\underset{q_{3}}{\sum}J_{-q_{1}+2q_{3}}(\zeta_{1})J_{-q_{3}}(\zeta_{3}).$
Consequently, the direct ATI transition term can be rewritten as
\begin{equation}
\label{Td2}
T_{d}\approx V_{e}^{-1/2}\left(\widetilde{U}_{p_{1}}+\widetilde{U}_{p_{2}}-q_1\omega_1-q_2\omega_2\right)
\Phi_{i}\left(\mathbf{p}_{f}\right)J_{-q_{1}}(\zeta_{1},\zeta_{3})J_{-q_{2}}(\zeta_{2}).
\end{equation}
This equation indicates that the entanglement terms between the two lasers (i.e., $J_{-q_{5}}(\zeta_{5})$ and $J_{-q_{6}}(\zeta_{6})$ in Eq.~(\ref{gbessel_linear}) can be ignored under our present atom-laser conditions. Thus the photon absorption from the two fields can be separately analyzed, where $J_{-q_{2}}(\zeta_{2})$ can be regarded as the probability that the atom absorbs $q_2$ photons from the $\omega_2$ laser field, while $J_{-q_{1}}(\zeta_{1},\zeta_{3})$ is the probability that the atom absorbs $q_1$ photons from the $\omega_1$ laser field. Moreover, we find that the photon number $q_2$ determines the height of the corresponding plateau to be proportional to $|J_{-q_{2}}(\zeta_{2})|^2$ in the terrace structure of the ATI spectrum, while the width of each plateau is determined by the photon number of $q_1$. We will explain these results separately below.

We now consider the height of each plateau in the terrace structure of the ATI spectrum. Under the present laser-atom conditions, we find that the absolute value of the argument $\zeta_2$ is much smaller than one, which causes the consistent decreasing of the value of the Bessel function  $J_{-q_{2}}(\zeta_{2})$ as the order $q_2$ increases. Therefore, it is reasonable to  define ATI channels by the photon number $q_2$ absorbed by the atom from the XUV laser field, such as the first channel is for $q_2=1$ and the second channel for $q_2=2$, etc..  Fig.~\ref{f1} is the ATI spectra we calculated for different channels. It can be found that each channel spectrum agrees well with the corresponding level in the terrace-like structure of the total ATI spectrum, indicating that the photon number of the XUV laser absorbed by the atom determines the height of the corresponding plateau in the terrace of the ATI spectrum.

Furthermore, we consider the width of each plateau in the spectrum for each ATI channel.  In order to understand what determines the width, we now focus on the analysis of the generalized Bessel function $J_{-q_{1}}(\zeta_{1},\zeta_{3})$ for a certain channel $q_2$. This Bessel function can be expressed in the form~\cite{wang1,guoyc,wang2}
\begin{align}
\label{bessel-q1}
J_{-q_{1}}\left(\zeta_{1},\zeta_{3}\right)=\frac{1}{T}\int_{-T/2}^{T/2}dt\exp\left\{ i\left[\zeta_{1}\sin\left(\omega_{1}t\right)+\zeta_{3}\sin\left(2\omega_{1}t\right)+q_{1}\omega_{1}t\right]\right\},
\end{align}
where $T=\frac{2\pi}{\omega_1}$. On the other hand, the classical action of an electron in a laser field is
\begin{align}
\label{s}
S_{c1}\left(\mathbf{p}_f,t\right)= & \frac{1}{2}\int_0^t dt'\left[\mathbf{p}_f-e\mathbf{A}_{c1}\left(t'\right)\right]^{2}\nonumber\\
= & \left(\frac{\mathbf{p}_f^{2}}{2}+\widetilde{U}_{p_{1}}\right)t-2\frac{\sqrt{u_{p_{1}}}}{\omega}|\mathbf{p}_f
\cdot\hat{\epsilon}|\sin\left(\omega t\right)+\frac{\widetilde{U}_{p_{1}}}{2}\sin\left(2\omega t\right),
\end{align}
where $\widetilde{U}_{p_{1}}=E_0^2/(4\omega_{1}^2)$ is the ponderomotive energy of the laser, and  $\textbf{A}_{c1}(t)=\hat{\epsilon} E_0/\omega_1 \cos(\omega_1 t)$ is the laser's vector potential with $E_0$ being the amplitude of the laser's electric field and $\hat{\epsilon}$ the direction of the laser polarization.
Comparing Eq.~(\ref{bessel-q1}) with Eq.~(\ref{s}), the Bessel function can be rewritten as
\begin{align}
\label{b2}
J_{-q_{1}}\left(\zeta_{1},\zeta_{3}\right)=  \frac{1}{T}\int_{-T/2}^{T/2}dt\exp \{i[S_{c1}\left(\mathbf{p}_f,t\right)-(q_2\omega_2-I_p) t]\}.
\end{align}
To obtain the above equation, we have applied the energy conversation of the initial and final states during the transition process, and ignored the $\widetilde{U}_{p_2}$ term because it is much smaller than one.

By using the saddle-point approximation, Eq.~(\ref{b2}) becomes $J_{-q_{1}}\left(\zeta_{1},\zeta_{3}\right)= \omega e^{if\left(t_{0}\right)}\sqrt{\frac{1}{2\pi if^{\prime\prime}\left(t_{0}\right)}}$ with $f(t)=S_{c1}\left(t,\mathbf{p}_f\right)-\left(q_{2}\omega_2-I_{p}\right)t$.
The saddle point $t_0$ satisfies $f'(t)=0$, which leads to the important energy relationship
\begin{align}
\label{energy}
\frac{\left[\mathbf{p}_{f}-e\mathbf{A}_{cl}\left(t_{0}\right)\right]^{2}}{2}=q_{2}\omega_{2}-I_{p},
\end{align}
where this equation expresses the energy conservation when the electron at time $t_0$ is ionized into the continuum from the bound state by absorbing $q_{2}$ photons of the XUV laser field.

Based on the above equation, the cutoff value of the kinetic energy for channel $q_2$ can be easily obtained as $E_{\rm max}=[\sqrt{2(q_2\omega_2-I_p))}+E_0/\omega_1]^2/2$ with $t_0=(2n+1)\pi/\omega_1$ for $n$ being an integer. The minimum value of the kinetic energy of channel $q_2$ can be classified into two cases: if the value of $q_2\omega_2-I_p$ is smaller than $2\widetilde{U}_{p_1}$, then the minimum kinetic energy is zero; otherwise it is $E_{\rm min}=[\sqrt{2(q_2\omega_2-I_p)}-E_0/\omega_1]^2/2$. Therefore, the beginning and the cutoff of the ATI spectrum for each channel can be determined by Eq.~(\ref{energy}). The arrows in Fig.~\ref{f1} show the beginning and cutoff positions of each plateau in the terrace structure, which indicates that the numerical results agree very well with the prediction of Eq.~(\ref{energy}). Figure 2 presents the schema of the transition process for the first channel (a), the second channel (b), and the third channel (c), according to Fig.~\ref{f1}.

It is quite interesting to understand the multiphoton transition process in terms of a classical field viewpoint: it tells us that, although in a quantum multiphoton transition process the electron absorbs many photons from both laser fields, Eq.~(\ref{Td2}) indicates that, when the electron absorbs $q_2$ photons at time $t_0$ from the high-frequency laser, it is ``exposed" to the low-frequency laser field at this time, and hence it can be accelerated (decelerated) by that field and obtains higher (lower) kinetic energy. Because $t_0$ can be at any time in one laser cycle, the final kinetic energy of the ionized electron changes between the minimum and maximum values determined by Eq.~(\ref{energy}).
Consequently, the ATI spectrum presents a wide plateau for each $q_2$ channel, resulting in a terrace-like structure in the final ATI spectrum when the contributions from all channels are considered. Figure~\ref{f3} presents the ATI spectra as a function of the intensity of the two-color laser field. We may find that the heights of the plateaus in the ATI spectra increase with the intensity of the XUV laser ( Fig.~\ref{f3}(a)), and the widths of the plateaus increase with the intensity of the IR laser, where the cutoffs can still be predicted by Eq.~(\ref{energy}) (Fig.~\ref{f3}(b)). These results confirm the roles of the two color laser fields in an ATI process.

At present, the free-electron laser technology can provide a laser field with very high frequency, where one photon energy can be much larger than the atomic ionization threshold. In this work, we find that the difference of the heights between two neighboring plateaus in the ATI spectrum increases with the frequency of the XUV laser, which can be seen in Fig.~\ref{f4}. This result can be understood easily based on our frequency-domain theory: because one photon energy of the XUV laser field is much larger than the atomic ionization threshold, the probability for the multiphoton transition decreases very rapidly with the absorbed XUV photon number. Consequently, the ionization probability  by the XUV laser decreases more rapidly as the XUV laser frequency increases, and hence only single-photon absorbing channel provides a dominant contribution to the transition process as the frequency is very large, as shown in Fig.~\ref{f4} (d).

Figure 5 shows the ATI spectra versus XUV photon energy $\omega_2$ and the final photoelectron kinetic energy. In Fig.~\ref{f5}, the black curve predicts the beginning position of the first plateau, while the violet and magenta curves describe the cutoff positions for the first and second plateaus, respectively. One may find that these curves agree well with the quantum numerical results. Moveover, we can see that the widths of all plateaus increase with $\omega_2$, while the transition rates of all plateaus decrease with $\omega_2$. Consequently, when $\omega_2$ is larger than 60$\omega_1$ in our computational conditions, only one plateau is present.

\section{conclusions}

By using the frequency-domain theory, we have investigated the ionization process of an atom in an IR+XUV two-color laser field where the XUV photon is close to or larger than the atomic ionization threshold. By analyzing the contributions from each high-frequency photon channel, we have found that the two laser fields play different roles in the ionization process: the XUV laser field determines the probability of the ionization while the IR laser field determines the final kinetic energy of the ionized electron. Furthermore, as the frequency of the XUV laser increases, the difference of the ionization probability between the neighboring XUV-photon channels increases. Therefore, only the first XUV-photon channel provides a dominant contribution in the ATI process as the XUV photon energy is much larger than the atomic ionization threshold, and only one plateau can be seen in the ATI spectrum.

\begin{acknowledgments}
BW thanks the members of the SFAMP club for helpful discussions and suggestions. This research was supported by the National Natural Science Foundation of China
under Grant No. 11074296, No. 61275128 and No. 11274050. ZCY was supported by NSERC of Canada and by the Canadian computing facilities of
SHARCnet and ACEnet. JC was also supported by the National Basic Research Program of China Grant No. 2013CB922200. KZ was also partly supported by the National Basic Research Program of China Grant No. 2012CB921302.
\end{acknowledgments}

\begin{figure}
\includegraphics[width=14cm, height=10cm]{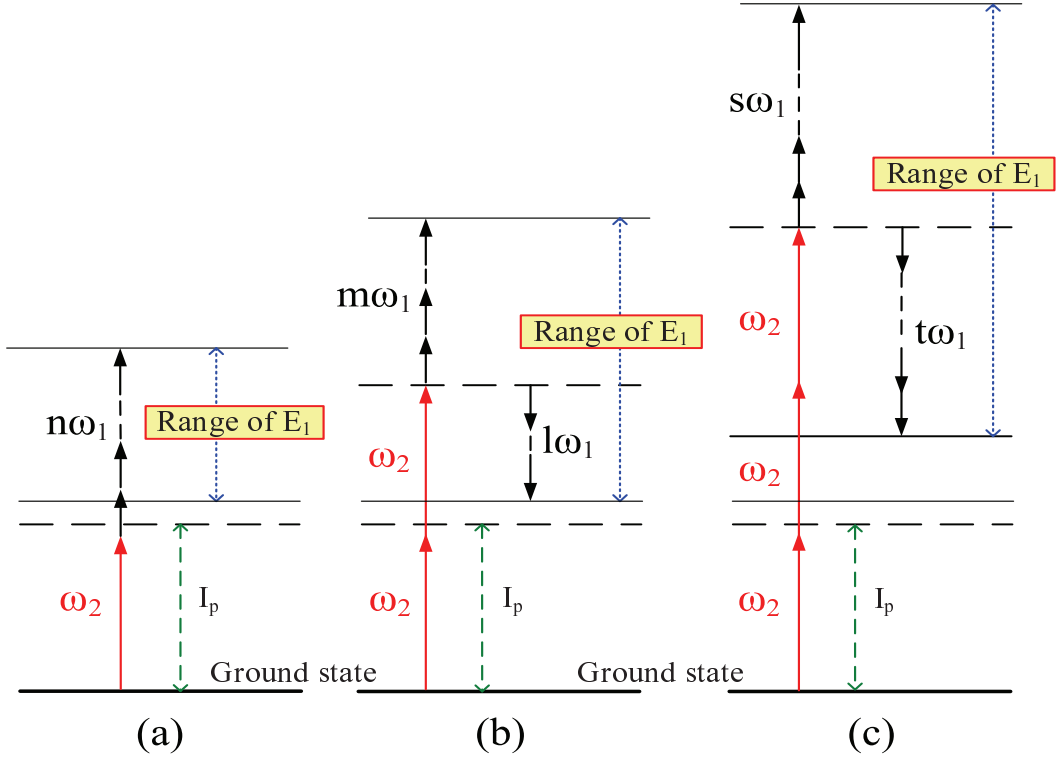}
\caption{(Color online) Multiphoton transition schema for (a) 1st channel, (b) 2nd channel, and (c) 3rd channel. $I_{p}$ is the ionization energy; $\omega_{1}$ and $\omega_{2}$ indicate the frequencies of IR and XUV laser fields, respectively; m,n,l,s,t are the number of IR photons absorbed or emitted by the electron. These three transition processes correspond to the first three plateaus in Fig.~\ref{f1}.}
\label{f2}
\end{figure}

\begin{figure}
\includegraphics[width=14cm, height=10cm]{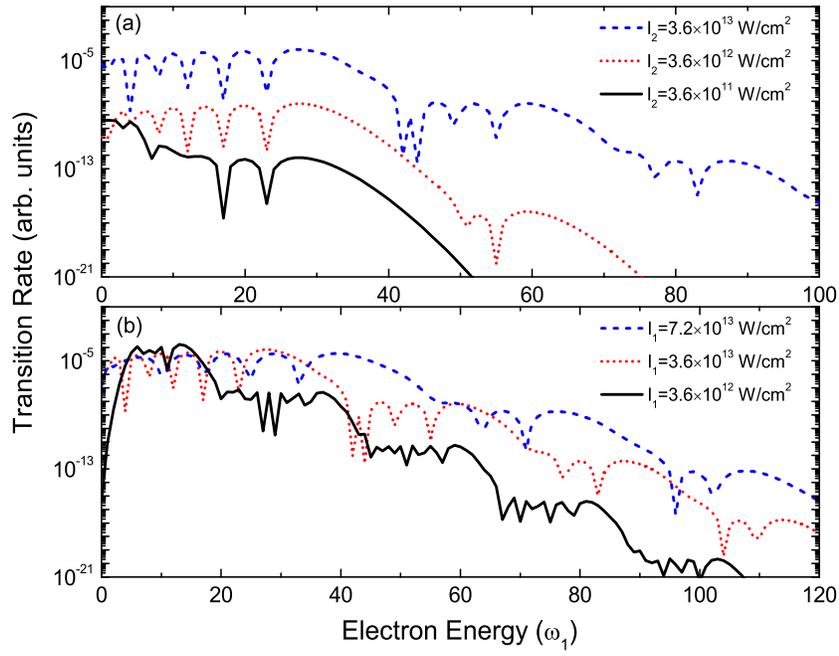}
\caption{(Color online) ATI spectrum for $\omega_2=20\omega_1$ with (a) laser intensity $I_1=3.6\times 10^{13}$ W/cm$^2$, and (b) laser intensity  $I_2=3.6\times 10^{13}$ W/cm$^2$.}
\label{f3}
\end{figure}

\begin{figure}
\includegraphics[width=14cm, height=10cm]{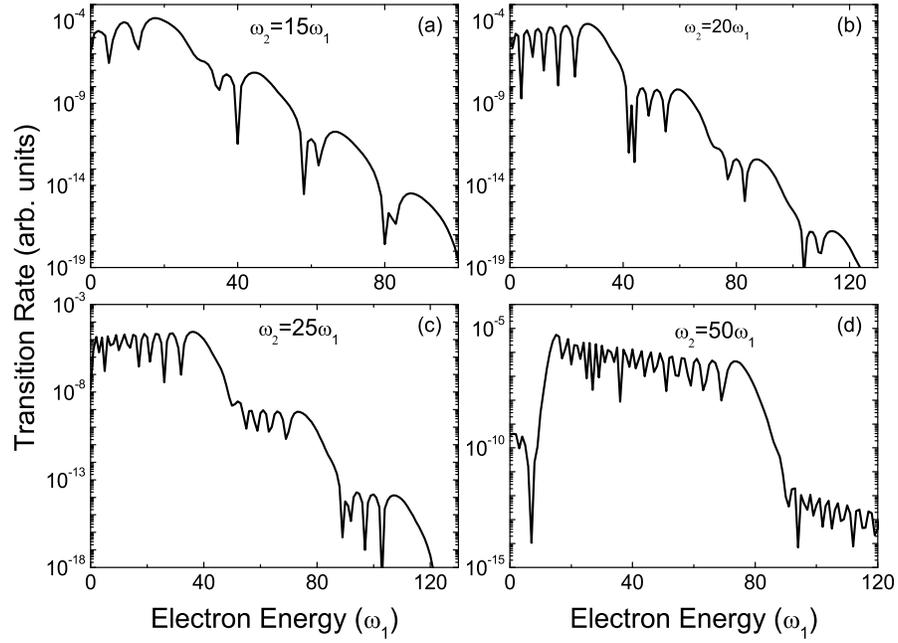}
\caption{(Color online) ATI spectrum for $\omega_2=15\omega_1$ (a), $\omega_2=20\omega_1$ (b), $\omega_2=25\omega_1$ (c), and $\omega_2=50\omega_1$ (d).}
\label{f4}
\end{figure}

\begin{figure}
\includegraphics[width=14cm, height=10cm]{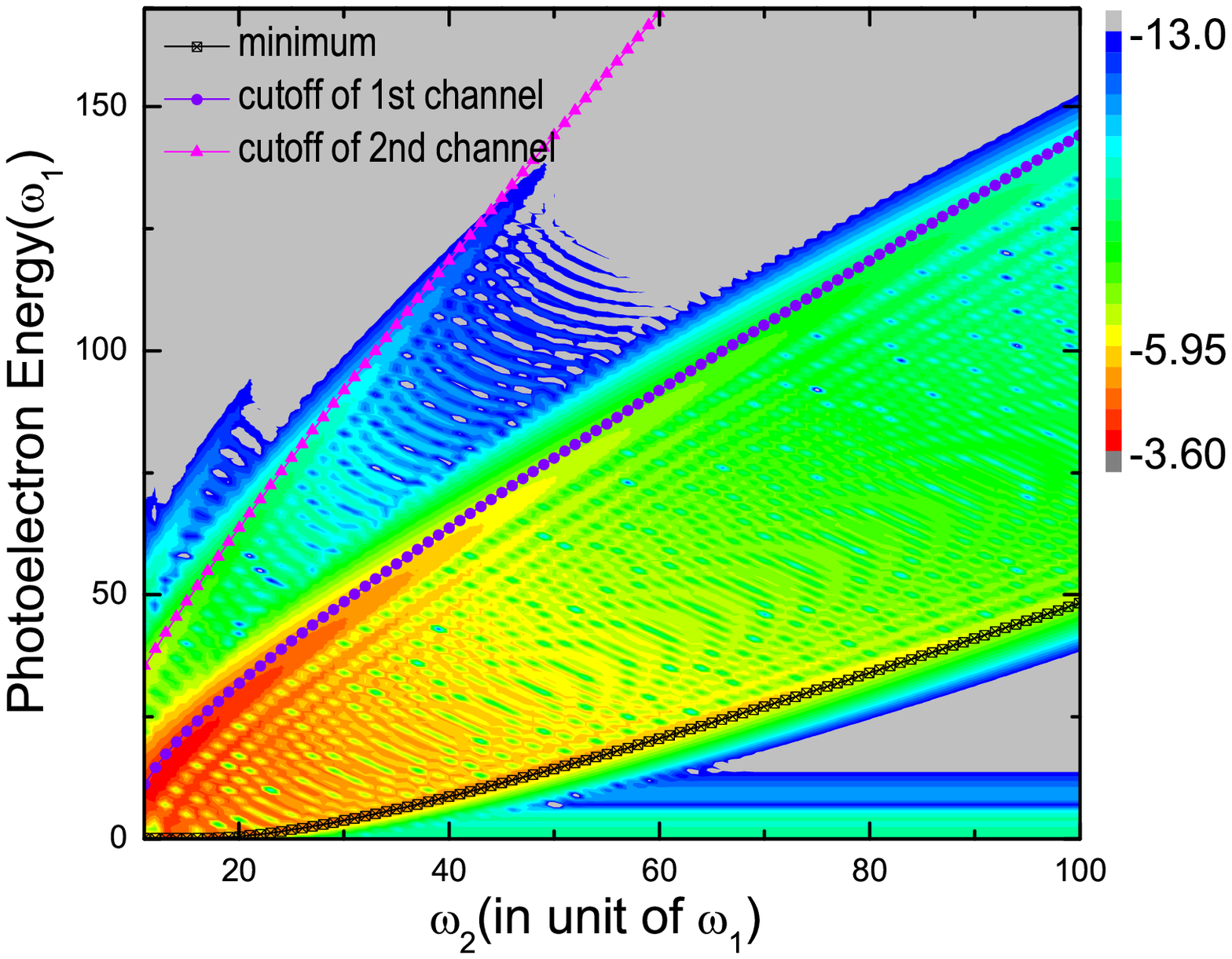}
\caption{(Color online) ATI spectrum versus XUV photon energy $\omega_2$ and the final photoelectron kinetic energy.
The black curve shows the beginning position of the first plateau, while the violet and magenta curves are the cutoff positions for the first and second plateaus. These curves are obtained by Eq.~(11), and they agree well with the numerical results.}
\label{f5}
\end{figure}


\begin{thebibliography}{apssamp}

\bibitem{a1}
P. Agostini, F. Fabre, G. Mainfray, G. Petite, and N. K. Rahman, Phys. Rev. Lett. {\bf42}, 1127 (1979).

\bibitem{hhg1}
A. McPherson et al., J. Opt. Soc. Am. B {\bf 4}, 595 (1987).

\bibitem{hhg2}
M. Ferray, A L'Huillier, X. F. Li, L. A. Lomprk, G. Mainfray and C. Manus, J. Phys. B {\bf 21}, L31 (1988).

\bibitem{hhg3}
T. Popmintchev, M.-C. Chen, P. Arpin, M. M. Murnane, and H. C. Kapteyn,  Nature Photonics {\bf 4}, 822 (2010).

\bibitem{fel1}
H. Motz, J. Appl. Phys. {\bf 22}, 527 (1951).

\bibitem{fel2}
B. W. J. McNeil and N. R. Thompson,  Nature Photonics {\bf 4}, 814 (2010).

\bibitem{schins}
J. M. Schins et al., Phys. Rev. Lett. {\bf73}, 2180 (1994).

\bibitem{tong}
N. Shivaram, H. Timmers, X.-M. Tong, and A. Sandhu, Chem. Phys. {\bf 414}, 139 (2013).

\bibitem{maquet}
V. Veniard, R. Taieb, and A. Maquet, Phys. Rev. Lett. {\bf74}, 4161 (1995).

\bibitem{fe}
C. Leone, S. Bivona, R. Burlon, and G. Ferrante, Phys. Rev. A {\bf 38,} 5642 (1988).

\bibitem{m}
D. B. Milosevic, and F. Ehlotzky, Phys. Rev. A {\bf 57,} 2859 (1998).

\bibitem{kk}
A. K. Kazansky, I. P. Sazhina, and N. M. Kabachnik, Phys. Rev. A {\bf 82}, 033420 (2010);
\textit{ibid}. {\bf86}, 033404 (2012);
J. Phys. B. {\bf 46}, 025601 (2013).

\bibitem{meyer}
P. Radcliffe et al., N. J. Phys. {\bf 14}, 043008 (2012).

\bibitem{guods}
D.-S. Guo and T. \AA berg, J. Phys. A {\bf  21}, 4577 (1988); D.-S. Guo, T. Aberg, and B. Crasemann, Phys. Rev. A {\bf40}, 4997 (1989).

\bibitem{gaolh}
L. Gao, X. Li, P. Fu, R. R. Freeman, and D.-S. Guo, Phys. Rev. A {\bf61,} 063407 (2000).

\bibitem{fu}
P. Fu, B. Wang, X. Li, and L. Gao, Phys. Rev. A {\bf 64,} 063401 (2001).

\bibitem{wang1}
B. Wang, L. Gao, X. Li, D.-S. Guo, and P. Fu, Phys. Rev. A {\bf 75}, 063419 (2007).

\bibitem{guoyc}
Y. Guo, P. Fu, Z.-C. Yan, J. Gong, and B. Wang, Phys. Rev. A {\bf 80}, 063408 (2009).

\bibitem{wang2}
B. Wang, Y. Guo, B. Zhang, Z. Zhao, Z. -C. Yan and P. Fu, Phys. Rev. A {\bf 82}, 043402 (2010).

\bibitem{wang3}
B. Wang, Y. Guo, J. Chen, Z.-C. Yan and P. Fu, Phys. Rev. A {\bf85}, 023402 (2012).

\bibitem{18}
M. Gell-Mann and M. L. Goldberger, Phys. Rev. {\bf 91}, 398 (1953); M. L. Goldberger and K.M. Watson, {\it Collision Theory} (Wiley, New York, 1964).

\bibitem{guo1992b}
D.-S. Guo and G. W. F. Drake, J. Phys. A \textbf{25}, 5377 (1992).



\end{thebibliography}
\end{document}